\documentclass[a4paper,12pt]{article}
\usepackage{amsmath,amssymb,epsfig}

\def\eqref#1{Eq.~(\ref{#1})}

\def\const{{\rm const}}

\def\bea{\begin{eqnarray}}
\def\eea{\end{eqnarray}}

\def\phi{\varphi}

\def\({\left(}
\def\){\right)}
\def\[{\left[}
\def\]{\right]}
\def\<{\left\langle}
\def\>{\right\rangle}

\def\la{\langle}
\def\ra{\rangle}
\def\bl{\bigl}
\def\br{\bigr}

\def\d{\partial}

\def\vk{{\bf k}}
\def\vu{{\bf u}}
\def\vB{{\bf B}}

\def\vb{{\skew{-4}\hat{\bf b}}}

\def\teq{t_0}
\def\tres{t_\eta}

\def\krms{k_{\rm rms}}

\def\kpar{k_{\parallel}}

\def\ls{\ell_{\rm s}}

\def\ld{\ell_\nu}
\def\lf{\ell_0}

\def\lres{\ell_{\eta}}

\def\Re{{\rm Re}}

\def\Pr{{\rm Pr}}

\def\Bsq{{\langle B^2 \rangle}}

\begin{document}

%{\footnotesize\noindent {\em Paper~P-4.035}\hfill
%Proceedings of the~29th~EPS Conference on Plasma Physics\\
%\null\hfill and Controlled Fusion (Montreux, 17-21~June~2002)
%\vskip0.5cm}

\centerline {\footnotesize\em 29th EPS Conf. on Plasma Phys. 
and Contr. Fusion, Montreux 17-21 June 2002, ECA~{\bf 26B}, 
P-4.035 (2002)}
\vskip0.5cm

\centerline{\rmfamily\bfseries\large THE NONLINEAR SMALL-SCALE DYNAMO AND} 
\centerline{\rmfamily\bfseries\large ISOTROPIC MHD TURBULENCE}

\vskip0.2cm

\centerline{A~A~Schekochihin,$^a$ 
S~C~Cowley,$^a$ G~W~Hammett,$^{a,b}$ J~L~Maron$^a$ \& J~C~McWilliams$^c$} 
%\centerline{and J~C~McWilliams$^c$}

\vskip0.1cm

{\footnotesize 
\centerline{\em $^a$Imperial College, London~SW7~2BW, England}
%\vskip-0.15cm
\centerline{\em $^b$Princeton Plasma Physics Laboratory, 
Princeton, New Jersey~08543, USA}
%\vskip-0.15cm
\centerline{\em $^c$UCLA, Los Angeles, California~90095-1565, USA}
}

\vskip0.2cm

Homogeneous incompressible 
MHD turbulence has been studied in the literature in two 
main regimes: with and without an externally imposed uniform 
mean magnetic field. In the former, explicitly anisotropic, case, 
Kolmogorov-style phenomenologies~\cite{Kraichnan,GS} predict 
a state of detailed scale-by-scale equipartion between 
hydrodynamic and magnetic fluctuations, which are Alfv\'en 
waves propagating along the mean field. With proposed scalings 
for the resulting spectra ranging between $k^{-2}$ and $k^{-5/3}$, 
both the hydrodynamic and the magnetic energies are concentrated 
at the outer (forcing) scale of the turbulence.  
It has long been the commonly held view that the other, 
{\em isotropic}, regime of MHD turbulence would, in the fully 
developed state, be very similar, with 
the energetic magnetic fluctuations at the outer scale
providing the effective mean field to support the Alfv\'en 
waves in the inertial range~\cite{Kraichnan}.
The isotropic regime is closely related to the 
problem of {\em turbulent MHD dynamo,} where a weak initial 
magnetic field is embedded into a turbulent conducting 
medium and its evolution is studied. In view of 
the arguments above, the dynamo has been expected 
to culminate in an Alfv\'enic equipartition state. 
In what follows, we will see this outcome is less obvious 
than it seems.

Because of the close correlation between velocities 
and magnetic fields in the Alfv\'en regime, the ratio 
between the fluid viscosity and the magnetic diffusivity 
(the magnetic Prandtl number~$\Pr=\nu/\eta$) 
has not been considered important, as long as both 
$\nu$ and $\eta$ were small enough for the turbulence to develop. 
However, it is a prominent property of 
hot low-density turbulent astrophysical plasmas (interstellar 
medium, some accretion discs and jets, protogalaxies, galaxy clusters, 
early Universe etc.) that their $\Pr$ tends to be extremely large 
and, therefore, that they possess a broad range of subviscous scales 
at which magnetic fields can exist, while velocities cannot.
The magnetic field lines are nearly perfectly frozen into 
such a highly conducting fluid. The fluid motions, even though 
restricted to scales above the viscous scale~$\ld$, can excite 
magnetic fluctuations at much smaller scales via stretching 
and folding of the field lines. 
In the kinematic (weak-field) regime, the result is 
an exponentially fast pile-up of magnetic energy at 
the resistive scale~$\lres\sim\Pr^{-1/2}\ld$, 
with a~$k^{3/2}$ spectrum extending through the subviscous 
range. The associated 
time scale is the turnover time of the viscous-scale turbulent 
eddies~\cite{Batchelor,KA}. 

The salient feature of the kinematically generated small-scale fields 
is their {\em folding structure:} the fields are ``folded'' in 
such a way that the smallness of their characteristic scale 
is due to rapid transverse spatial oscillation of the field 
direction, while the field lines remain largely unbent 
up to the scale of 
the stretching eddy~\cite{Ott,MCM_dynamo,SCMM_folding,SMCM_structure}.
Quantitatively, the field structure can be studied in terms 
of statistics of the field-line curvature~$K=|\vb\cdot\nabla\vb|$ 
(where~$\vb=\vB/B$) and of its correlation with the field strength~$B$.
In the kinematic case, analytic theory is possible subject 
to certain modelling assumptions~\cite{SCMM_folding}. 
The bulk of the PDF of~$K$ turns out to be at the velocity scales, 
with a power tail~$\sim K^{-13/7}$ extending through the 
subviscous range~$K\gg\ld^{-1}$ (this scaling is very well confirmed 
numerically~\cite{SCMM_folding}). This reflects the predominant straightness 
of the folded field lines at subviscous scales. 
They are significantly curved only 
in the ``corners'' of the folds --- the power tail of the curvature 
PDF describes the intermittent distribution of these corners. 
Furthermore, the field strength is anticorrelated with the field-line 
curvature, i.e., the field is stronger in the straight parts of 
the folds than in the corners~\cite{SCMM_folding,SMCM_structure}. 

The most important implication of these results is that 
the system is in a {\em reduced-tension state:} the tension 
force, which controls the nonlinear back reaction, is 
$\vB\cdot\nabla\vB\sim KB^2\sim B^2/\ld$ during the 
kinematic stage~\cite{SCMM_folding}, 
so the nonlinearity becomes important 
($\vB\cdot\nabla\vB\sim\vu\cdot\nabla\vu$) 
when the magnetic energy 
approaches that of the viscous-scale eddies (in a chaotically 
tangled field, this would happen at much smaller magnetic energies, 
because we would have~$\vB\cdot\nabla\vB\sim B^2/\lres$).
Even after the onset of the nonlinear regime, the field 
in the corners remains weak and cannot 
be expected to exert a significant amount of back reaction 
on the flow. In fact, the folding 
structure, once set up, is generally very hard to undo, as the detailed 
``unwinding flows'' required for that cannot exist at 
subviscous scales. These arguments are corroborated by the numerical 
simulations~\cite{SMCM_structure}, which show the persistence 
of the folding structure in the nonlinear regime. In particular, 
both the anticorrelation between the field strength and curvature 
and the~$K^{-13/7}$ subviscous-range scaling of the curvature PDF 
remain unchanged. The rigidity of the folding structure plays 
a key role in the further developments. 

We now propose the following scenario of the nonlinear-dynamo 
evolution~\cite{SCHMM_ssim}. 
The MHD induction equation gives the evolution law for the magnetic energy:
\bea
\label{W_eq}
\d_t\Bsq = 2\gamma(t)\Bsq - 2\eta\krms^2(t)\Bsq,\quad{\rm where}\quad 
\gamma(t) = {\la\vB\vB:\nabla\vu\ra/\Bsq}, 
\eea
and~$\krms(t)$ is the spectrum-integrated rms wave number 
of the magnetic field. In~\eqref{W_eq}, $\gamma(t)$ can be interpreted 
as the effective stretching rate at time~$t$. During the kinematic 
stage, $\gamma(t)\sim u_{\ld}/\ld$, the turnover rate of the 
viscous-scale eddies. 
Once the magnetic energy becomes comparable to the energy of these 
eddies, the Lorentz back reaction must 
act to suppress the stretching motions associated with them. 
However, the next-larger-scale eddies still 
have energies above that of the magnetic field, 
though the turnover rate of these eddies is smaller. 
These eddies will continue to amplify the field 
at this slower rate. 
The scale of these eddies is~$\ls>\ld$, so the folds 
are elongated accordingly and the tension force 
is~$\vB\cdot\nabla\vB\sim B^2/\ls$.
The corresponding inertial term is~$\vu\cdot\nabla\vu\sim u_{\ls}^2/\ls$, 
so these eddies become suppressed when~$B^2\sim u_{\ls}^2$, 
whereupon it will be the turn of the next-larger eddies to provide 
the dominant stretching action. 
Thus, at any given time, the ``stretching scale''~$\ls(t)$ is defined 
by $u^2_{\ls(t)}\sim\Bsq(t)$ and $\gamma(t)$~is the turnover 
rate of the eddies of scale~$\ls(t)$: 
$\gamma(t)\sim u_{\ls(t)}/\ls(t)$. The first term on the rhs 
of~\eqref{W_eq} is then 
$\gamma(t)\Bsq(t) \sim u^3_{\ls(t)}/\ls(t) \sim \epsilon$, 
where $\epsilon=\const$~is the Kolmogorov energy flux. 
The physical meaning of this result is as follows. 
The turbulent energy injected at the forcing scale cascades 
hydrodynamically down to the scale~$\ls(t)$ where 
(a finite fraction of) it is 
diverted into the small-scale magnetic fields. 
We conclude that the magnetic energy 
should grow linearly with time during this stage, $\Bsq(t)\sim\epsilon t$, 
and $\gamma(t)\sim1/t$. 
We stress that the magnetic field is still organized in folds of 
characteristic 
length~$\ls(t)$ with direction reversals at the resistive scale~$\lres$.
Comparing the two terms in the rhs of~\eqref{W_eq}, we can 
estimate 
$\lres\sim\krms^{-1}(t)\sim\[\gamma(t)/\eta\]^{-1/2}\sim\(\eta t\)^{1/2}$, 
so $\lres$ increases. Indeed, as the stretching slows down, 
{\em selective decay} eliminates the modes at the extreme UV end 
of the magnetic spectrum for which the resistive time is now shorter 
than the stretching time and which, consequently, cannot 
be sustained anymore. 

Some fluid motions do survive at scales 
below~$\ls(t)$ and down to the viscous cutoff. 
They are Alfv\'en waves that propagate along the 
folds of direction-reversing magnetic fields 
and do not amplify the magnetic energy~\cite{SCHMM_ssim}. 
Their dispersion relation is
$\omega=\pm\kpar\sqrt{\Bsq}$, where~$\kpar=\bl(\vk\vk:\vb\vb\br)^{1/2}$. 
A finite fraction of the hydrodynamic 
energy arriving from the large scales is channelled into 
the turbulence of these waves. 
Since $\ls^{-1}\ll\kpar\ll\ld^{-1}$, $\omega$~is 
larger than the resistive-dissipation rate of the small-scale 
fields: $\omega\gg u_{\ls}/\ls\sim\gamma\sim\eta\krms^2$. 
Therefore, the Alfv\'en waves are mostly dissipated {\em viscously} 
via the MHD turbulent cascade~\cite{Kraichnan,GS}, rather than resistively. 
This enables us to consider the evolution of the small-scale 
fields separately from that of the Alfv\'en waves. 
%as long as~$\lres\ll\ld$.

The nonlinear-growth stage 
continues until the magnetic energy 
becomes comparable to the energy of the outer-scale eddies, 
$\Bsq\sim u_{\lf}$, and $\ls\sim\lf\sim\Re^{3/4}\ld$, the forcing 
scale. Thus, energy equipartition between 
magnetic and velocity fields is achieved. 
The time scale for this to happen 
is the turnover time of the outer-scale 
eddies~$\teq\sim(u_{\lf}/\lf)^{-1}\sim\Re^{1/2}(u_{\ld}/\ld)^{-1}$. 
At this point,~$\gamma\sim u_{\lf}/\lf$. 
Therefore, the resistive scale is now~$\lres\sim\(\Re\,\Pr\)^{-1/2}\lf$,  
which is larger than its kinematic value~$\sim\Pr^{-1/2}\ld$ 
by a factor of~$\Re^{1/4}$.  
As there are no scales in the system larger than $\lf$, 
there can be no further growth of the magnetic energy. 
Stretching and bending the ever 
more rigid magnetic field becomes increasingly harder and, instead 
of amplifying the magnetic energy, causes the field to spring 
back. Therefore, the rate of the energy transfer 
into the small-scale magnetic 
field decreases, as this channel of energy dissipation 
becomes inefficient. 
Instead, we expect the energy injected by the forcing to be 
increasingly diverted into the Alfv\'enic turbulence 
that is left throughout the inertial range in the wake of the 
suppression of the stretching motions.  
During this last stage of the nonlinear dynamo, 
the magnetic energy stays approximately constant 
(growing only very slightly) while $\gamma(t)$ drops 
below the turnover rate of the outer-scale eddies. 
The energy balance~\eqref{W_eq} then implies further movement 
of $\lres\sim\krms^{-1}(t)$ towards larger scales. 
The mechanism for this decrease is the same as in the 
nonlinear-growth stage: the resistive decay of the high-$k$ 
modes outpaces their regeneration by the weakened stretching. 

Thus, $\lres$ increases both in the nonlinear-growth 
stage and during the subsequent slower approach to saturation. 
In~\cite{SCHMM_ssim}, we show that the evolution of the magnetic-energy 
spectrum in both regimes is likely to be {\em self-similar} 
with $\krms(t)\sim(\eta t)^{-1/2}$. 
It is very important to understand how far it 
can proceed. In our arguments so far, we have disregarded 
the Alfv\'enic component of the turbulence. 
It is not, however, justified to do so once the small-scale 
magnetic energy reaches the velocity scales. 
Indeed, the decrease of~$\krms$ is basically a consequence 
of the balance between the field-amplification 
and the resistive-dissipation terms in~\eqref{W_eq}. 
However, once $\krms\sim\ld^{-1}$, 
the Alfv\'enic turbulence will start to affect~$\krms$ 
in an essential way: since the waves are damped 
at the viscous scale, $\krms$~must stabilize at $\krms\sim\ld^{-1}$.  
The resulting turbulent state features folded 
magnetic fields reversing directions at the viscous scale plus 
Alfv\'en waves in the inertial range propagating along the folds. 
There are then two possibilities: either (i)~this represents the 
final steady state of the isotropic MHD turbulence, 
or (ii)~further evolution will lead to unwinding of the 
folds and continued energy transfer to larger scales, 
so that the spectrum will eventually peak at the outer scale and 
have an Alfv\'enic power tail extending through the inertial 
range. It is in the latter case that 
the turbulence in the inertial range would 
be of the usual Alfv\'en-wave kind~\cite{Kraichnan,GS}, where 
the large-scale magnetic fluctuations would provide a mean field along which 
the inertial-range Alfv\'en waves propagate. 
However, in order to achieve such a state, the folds 
would have to be unwound. In view of the rigidity of the folding 
structure, it is unclear how this can be done. 
This dichotomy remains unresolved and requires further study. 

In either case, $\krms\sim\ld^{-1}$, so the fully developed isotropic 
MHD turbulence is characterized by the equalization of the 
resistive and viscous scales. 
The time scale at which such an equalization is brought about 
is the resistive time associated with the viscous 
scale of the turbulence:~$\tres(\ld)\sim\ld^2/\eta$. 
We immediately notice that, in order for the nonlinear-growth stage 
to run its full course up to the energy equipartion~$\Bsq\sim u_{\lf}^2$,  
this time scale must be longer than the turnover time of 
the outer-scale eddies: $\tres(\ld)\gg\teq$, which 
requires~$\Pr\gg\Re^{1/2}$ and is also equivalent to 
the condition that~$\lres\ll\ld$ 
at the end of the nonlinear-growth stage. 
This constitutes the ``true large-$\Pr$ regime,'' which is 
the one relevant for astrophysical plasmas. 
In this regime, the magnetic energy saturates at the equipartition 
level, $\Bsq\sim u_{\lf}^2$. From~\eqref{W_eq}, 
we get an estimate of the amount of turbulent power that is still dissipated 
resisively: $\gamma\Bsq \sim (\eta/\ld^2)u_{\lf}^2 
\sim \epsilon\,\Re^{1/2}/\Pr\ll\epsilon$, i.e., most 
of the injected power now goes into the viscously-dissipated 
Alfv\'enic motions. The other possibility is $\Pr\lesssim\Re^{1/2}$. 
In this case, the self-similar dynamo evolution is 
curtailed during the nonlinear-growth stage 
with magnetic energy still at a subequipartition value, 
$\Bsq/u_{\lf}^2 \sim \tres(\ld)/\teq \sim \Pr/\Re^{1/2}$,
and resistivity continuing to take a significant 
part in the dissipation of the turbulent energy. 
If no further evolution takes place, 
this gives an estimate of the saturation energy of 
the magnetic component of the turbulence. 
In this regime, $\Pr$ is not large enough 
to capture all of the physics of the large-$\Pr$ dynamo. 
Most of the extant numerical simulations appear to 
be in this regime (see references in~\cite{SCHMM_ssim}).

Thus, we have identified 
two possibilities for the long-time behaviour of 
the isotropic MHD turbulence: saturation 
in the usual Alfv\'enic state~\cite{Kraichnan,GS} 
and saturation with the magnetic energy 
tied up in the viscous-scale fields. Which one is realized 
depends on the way the small-scale folded fields interact 
with the inertial-range Alfv\'enic turbulence. 
We stress that there is no numerical evidence 
available at present that would confirm that, the isotropic 
MHD turbulence {\em without externally imposed mean field} 
--- at any Prandtl number --- 
attains the Alfv\'enic state of scale-by-scale equipartition envisioned 
in the commonly accepted phenomenologies~\cite{Kraichnan,GS}. 
In fact, medium-resolution simulations~\cite{MCM_dynamo} 
rather seem to support the final states with small-scale 
energy concentration even for~$\Pr=1$. 
This does not mean that the Alfv\'enic picture is incorrect {\em per se}.
However, all existing phenomenologies of the Alfv\'enic turbulence 
depend on the assumption~\cite{Kraichnan} 
that the strongest magnetic fields are those at the outer scale. 
This is authomatically satisfied if a finite mean field is imposed 
externally. However, it remains to be seen if such a distribution of energy 
is set up self-consistently when the turbulence is isotropic.

\end{document}